\documentstyle[12pt,a4wide,epsfig]{article}

\def\D{\displaystyle}
\def\T{\textstyle}

\def\SS{\scriptscriptstyle}

\begin{document}
\thispagestyle{empty}
\noindent
\begin{flushright}
\normalsize{DO--TH 96/15\\}
\end{flushright}
\vspace*{5.0cm}
\begin{center}
{\Large
\boldmath $\varepsilon '/\varepsilon $ \unboldmath \bf COMPUTATION -- 
\\[0.3cm]
LONG-DISTANCE EVOLUTION}
\footnote{Invited talk presented at the `Workshop on K Physics', Orsay,
France, May 30 -- June 4, 1996. A slightly shortened version will 
be published in the proceedings.} \\[24pt] 
Peter H. Soldan \\
Institut f\"ur Physik, Universit\"at Dortmund \\
D-44221 Dortmund, Germany \\ 
\vspace*{5cm}
{\bf Abstract\\}
\end{center}

{\small
We present a status report on our study of long-distance
contributions to the decay ampli\-tudes $A(K^0\rightarrow 2\pi,\,I)$ 
in the framework of the $1/N$ expansion. 
We argue that a modified prescription for the
identification of meson momenta in the chiral loop corrections has to be
used to gain a self-consistent picture which allows an appropriate
matching with the short-distance part. Possible uncertainties in the analysis
of the density-density operators $Q_6$ and $Q_8$ which dominate
the CP violation parameter $\varepsilon '/\varepsilon$ are discussed.
As a first result we present the long-distance $1/N$ correction
to the gluon penguin operator $Q_6$ in the chiral limit.
}
\newpage
\section{Introduction}
The origin of CP asymmetries is an open issue within the electroweak theory.
In the standard model CP violation arises from a complex phase introduced by
hand in the Yukawa couplings. It is of special interest to investigate 
whether in this approach the ratio $\varepsilon '/\varepsilon$, the measure 
of direct CP violation in $K\rightarrow \pi\pi$ decays, may be 
close to zero, therefore mimicking the superweak model of Wolfenstein
\cite{Wolf}.

At present the experimental evidence is inconclusive,
\begin{equation}
Re\left(\frac{\varepsilon '}{\varepsilon}\right)=\left\{ 
\begin{array}{ll}
(23\pm 3.6\pm 5.4)\hspace{1mm}\cdot 10^{-4}& \hspace*{1.5cm} 
\mbox{NA 31 }\cite{NA31} \\
(7.4\pm 5.2\pm 2.9)\cdot 10^{-4} & \hspace*{1.5cm} \mbox{E 731 } \cite{E731}
\end{array} \right.\;, 
\end{equation} 
and the new experiments are awaited to provide more accurate data.

In the standard model the calculation of $\varepsilon '/\varepsilon$ is
based on the effective low-energy hamiltonian for $\Delta S=\nolinebreak 1$ 
transitions \cite{Mar},
\begin{equation}
{\cal H}_{ef\hspace{-0.5mm}f}^{\SS \Delta S=1}=\frac{G_F}{\sqrt{2}}
\;\xi_u\sum_{i=1}^8 c_i(\mu)Q_i(\mu)\;,
\label{ham}
\end{equation}
\begin{equation}
c_i(\mu)=z_i(\mu)+\tau y_i(\mu)\;,\hspace*{1cm}\tau=-\xi_t/\xi_u\;,
\hspace*{1cm}\xi_q=V_{qs}^*V_{qd}^{}\;, 
\end{equation}
where the Wilson coefficient functions $c_i(\mu)$ of the local four-fermion
operators $Q_i(\mu)$ are obtained by means of the renormalization group
equation. They were computed in a comprehensive next-to-leading order
analysis by two groups \cite{BJL,CFMR}. Using eq.~(\ref{ham}) the ratio 
$\varepsilon '/\varepsilon$ reads
\begin{equation}
\frac{\varepsilon'}{\varepsilon}=\frac{G_F}{2}\frac{\omega}{|\varepsilon|
|A_0|}Im\,\xi_t\sum_{i=1}^8 y_i \Big(\langle Q_i\rangle_0-\frac{1}{\omega}
\langle Q_i \rangle_2\Big)\; ,
\hspace*{1.3cm}
\omega^{-1}=\frac{Re\,A_0}{Re\,A_2}\simeq 22\;.
\label{eps}
\end{equation}
$\langle Q_i\rangle_I$ are the hadronic matrix elements for the isospin
states $I$ which contain the long-distance contribution to the amplitudes
$A_I$ for the process under consideration,
\begin{equation}
\langle Q_i(\mu)\rangle_I \equiv \langle \pi\pi, \,I|Q_i(\mu)|K^0\rangle\;.
\end{equation}
The dominant CP violating effects result from the gluon and the electroweak
penguins, $\langle Q_6\rangle_0$ and $\langle Q_8\rangle_2$ respectively,
with
\begin{equation}
Q_6 =-2\sum_{q=u,d,s}\bar{s}(1+\gamma_5) q\,\bar{q}(1-\gamma_5) d \;,
\hspace{1cm}   
Q_8=-3\sum_{q=u,d,s}e_q\,\bar{s}(1+\gamma_5) q\,\bar{q}(1-\gamma_5) d\;,
\end{equation}
where $\langle Q_6\rangle_2$ in addition arises through isospin breaking
effects ($\pi\eta\eta '$ mixing). Consequently, the main task within the
investigation  of long-distance effects is to estimate the degree of 
cancellation between the two penguin contributions in eq.~(\ref{eps}) which 
gives rise to a small value of the ratio $\varepsilon'/\varepsilon$.

Although recently there has been considerable progress in lattice
calculations of the hadronic matrix elements \cite{Lat}, it is still 
meaningful to improve the existing phenomenological studies in order to
see whether the results are in agreement.
There are several approximations of non-perturbative QCD available
\cite{BBG,SR,Fras}.
We shall perform our analysis applying the $1/N$ expansion
($N$ being the number of colors) where we will point 
out that a consistent matching between short- and long-distance contributions 
is possible, provided a proper identification of meson momenta in the chiral 
loop corrections is carried out. 
\section{General Framework: \boldmath $1/N$ \unboldmath Expansion}
To estimate the hadronic matrix elements we shall use the chiral effective
lagrangian for $K\rightarrow \nolinebreak \pi\pi$ decays, which involves an 
expansion in
meson momenta where terms up to ${\cal O}(p^2/\Lambda_\chi^2)$ are included,
$\Lambda_\chi$ being the chiral symmetry breaking scale of ${\cal O}(1GeV)$
\cite{BBG},
\begin{equation}
{\cal L}_{ef\hspace{-0.5mm}f}=\frac{f^2}{8}\mbox{Tr}\Big(D_\mu U 
D^{\mu}U^\dagger+r(m U^\dagger+Um^\dagger)\Big)-\frac{f^2 r}{8 
\Lambda^2_{\chi}}\mbox{Tr}\Big(m D^2 U^\dagger+D^2 Um^\dagger\Big)
\;.\label{Leff}
\end{equation}
$f$ and $r$ are free parameters related to the pion decay constant
$F_\pi$ and to the quark masses, respectively, $m=diag(m_u,\,m_d,\,m_s)$
being the quark mass matrix. The degrees of freedom of the complex matrix $U$ 
are identified with the pseudoscalar meson nonet, given in a non-linear
representation, 
\begin{equation}
U=\T\exp\frac{2i}{f}\pi^a\lambda_a = \exp\frac{i\sqrt{2}}{f}\left[
\begin{array}{ccc}
\pi^0+\frac{\T\eta_8}{\sqrt{3}}+\sqrt{\frac{2}{3}}\eta_0 & \sqrt2\pi^+ 
& \sqrt2 K^+  \\
\sqrt2 \pi^- & -\pi^0+\frac{\T\eta_8}{\sqrt{3}}+\sqrt{\frac{2}{3}}\eta_0 
& \sqrt2 K^0 \\
\sqrt2 K^- & \sqrt2 \bar{K}^0 & -\frac{2{\T\eta_8}}{\sqrt{3}}
+\sqrt{\frac{2}{3}}\eta_0    
\end{array} \right]\;.
\end{equation}
A straightforward bosonization yields the chiral representation of the
corresponding quark currents and densities, which allows to express the
four-fermion operators $Q_i$ in terms of the meson fields,
\begin{equation}
\begin{array}{lclcl}
(J^\mu_L)_{ij}&\equiv&\bar{q}_i\gamma^\mu(1-\gamma_5) q_j&=&
\D\frac{if^2}{4}\Big(2\partial^\mu
U U^\dagger -\frac{r}{\Lambda_\chi^2}(m\partial^\mu U^\dagger -\partial^\mu U 
m^\dagger)\Big)_{ji}\;,
\\[2mm]
(D_L)_{ij}&\equiv&\bar{q}_i(1-\gamma_5) q_j&=&
\D-\frac{f^2}{4}r\Big(U-\frac{1}{\Lambda_{\chi}^2}
\partial^2 U\Big)_{ji}\;.
\end{array}
\label{CD}
\end{equation}
The $1/N$ corrections to the matrix elements $\langle Q_i\rangle_I$ are
calculated by chiral loop diagrams. In these diagrams we encounter 
divergences which are regularized by a finite cutoff $\Lambda_{cut}$ as it
was introduced by Bardeen, Buras and G{\'erard} (BBG) to estimate the
long-distance contribution to the $\Delta I=1/2$ rule \cite{BBG}.

The inclusion of the one-loop corrections leads to a quadratic as well as to 
a logarithmic dependence of the matrix elements on the cutoff, i.e., the 
results are given in terms of 
\begin{equation}
\sim\frac{\Lambda_{cut}^2}{(4\pi f)^2}\;,\hspace{1cm}
\sim\frac{m_A^2}{(4\pi f)^2}
\ln\Big(1+\frac{\Lambda^2_{cut}}{m_B^2}\Big)
\end{equation}
($m_{A,\,B}$ being pseudoscalar meson masses), where the quadratic part
corresponds to the chiral limit ($m_q=0$). Actually, the loop expansion
involves a series in $1/f^2\sim1/N$, which is in direct accordance to
the short-distance expansion in $\alpha_s/\pi\sim 1/N$.

Due to the truncation to pseudoscalar mesons, $\Lambda_{cut}$ has to be taken
at or even below the ${\cal O}(1GeV)$. This restriction is a common feature
of the phenomenological approaches at hand, in which higher resonances are
not included. 

To retain the physical amplitudes $A_I$, which as a matter of principle are 
scale-inde\-pendent, the long- {\it and} the short-distance contributions are 
evaluated at the cutoff scale, i.e., the long-distance ultraviolet cutoff is 
identified with the short-distance infrared one. 
This procedure for the matching of the Wilson coefficient functions with the
hadronic matrix elements we shall analyse in detail in the following section.   
\section{Chiral Loop Corrections\label{CLC}}

As mentioned above the calculation of chiral loop effects motivated by
the $1/N$ expansion was introduced by BBG to explain the $\Delta I=1/2$ rule. 
The authors considered the loop corrections to the current-current operators
$Q_1$ and $Q_2$, whereas the gluon penguin operator $Q_6$ was included at the
tree level \cite{BBG}, implicitly assuming that the $1/N$ corrections 
to the latter are small.

Following the same lines Buchalla {\it et al.} \cite{Buch} performed a 
detailed analysis of the ratio $\varepsilon '/\varepsilon$, in which they
stressed that through the quadratic dependence of $\langle Q_6\rangle$
and $\langle Q_8 \rangle$ on the running mass $m_s$ the evolution of
the corresponding Wilson-coefficients is cancelled in the large-$N$ limit 
without considering chiral loops. One crucial result of their 
investigations is a small value of $\varepsilon'/\varepsilon$, caused
by a large cancellation between the gluon and the electroweak penguin 
contributions.

Heinrich {\it et al.} extended the analysis by including chiral
loops, i.e., the $1/N$ corrections to the matrix elements of $Q_6$ and $Q_8$
\cite{JMS1,JMS2}. As a net effect they reported an enhancement of 
$\langle Q_6\rangle_0$ and a large decrease for $\langle Q_8\rangle_2$,
through which the cancellation gets less effective. Numerically the authors
found a large positive value of $\varepsilon '/\varepsilon$ \cite{EAP2},
\begin{equation}
\varepsilon '/\varepsilon=(9.9\pm4.1)\cdot 10^{-4}\hspace{1.5cm}
[m_s(1GeV)=175\,MeV]\;.
\label{eap}
\end{equation}
The relative size of the $1/N$ corrections contained in eq.~(\ref{eap}) has to 
be compared with the factors $B_6$ and $B_8$, denoting deviations from the 
lowest order results for the corresponding matrix elements, as obtained e.g.~by 
lattice calculations. The latter give \cite{Lat}
\begin{equation}
B_6^{\small (I=0)}=1.0\pm 0.2\hspace{0.5cm}\mbox{and} \hspace{0.5cm}
B_8^{\small (I=2)}=1.0\pm 0.2\
\end{equation} 
(in accordance with a recent QCD sum rule analysis of $\langle Q_6 \rangle$ 
\cite{SR}), predicting again a strong cancellation 
between the various penguin contributions. 
In view of this different results we perform a detailed reconsideration of 
chiral loop effects in the $1/N$ approach.\\[12pt]
One crucial point within our analysis of the $K\rightarrow \pi\pi$
decay amplitudes is the matching of short- and long-distance contributions at 
the cutoff scale. In the {\it standard approach} referred to above, 
the effective color singlet gauge boson connecting 
the two currents and densities, respectively, is integrated out from
the beginning. The basic diagrams hence to be calculated are presented 
in fig.~1 where the squares denote weak vertices corresponding to the bosonized
four-fermion operators $Q_i$ and the circles indicate strong interaction
vertices.\\[12pt]
\hspace*{2.98cm}\epsfig{file=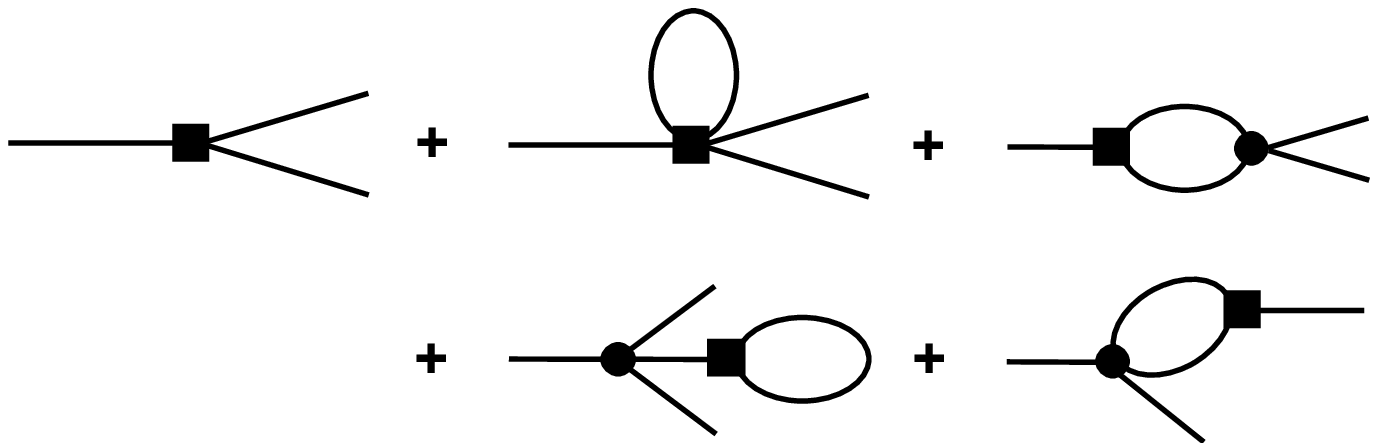,height=3.5cm}\\[12pt]
\centerline{\small Fig.~1: Chiral diagrams for $K\rightarrow\pi\pi$ decays.}
\\[12pt]
Within this procedure the cutoff $\Lambda_{cut}$ is associated to the meson
in the loop, i.e., the integration variable is identified with the meson
momentum. Consequently, as there is no corresponding quantity in the 
short-distance part, {\it a rigorous matching of long- and short-distance
contributions is not possible}.

The ambiguity is removed by associating the cutoff to the effective color
singlet gauge boson as introduced within a study of the $K_L$-$K_S$ mass 
difference \cite{BGK} and recently used by Fatelo and G{\'erard} who 
calculated the long-distance evolution of {\bf current-current operators} 
in the chiral limit applying the $1/N$ expansion \cite{FG}. 
The corresponding identification of the loop integration variable
with the momentum flow between the two currents and densities, respectively,
is feasible in the long- {\it and} the short-distance part of the
analysis, through which a direct momentum matching is justified.

The crucial feature of this {\it modified approach} is illustrated in fig.~2. 
The momentum of the virtual meson is shifted by the external momentum, 
the former being no longer identical to the integration variable
$q$, which affects both, the quadratic and the logarithmic behaviour of the
$1/N$ corrections.\\[12pt]
\hspace*{4.0cm}\epsfig{file=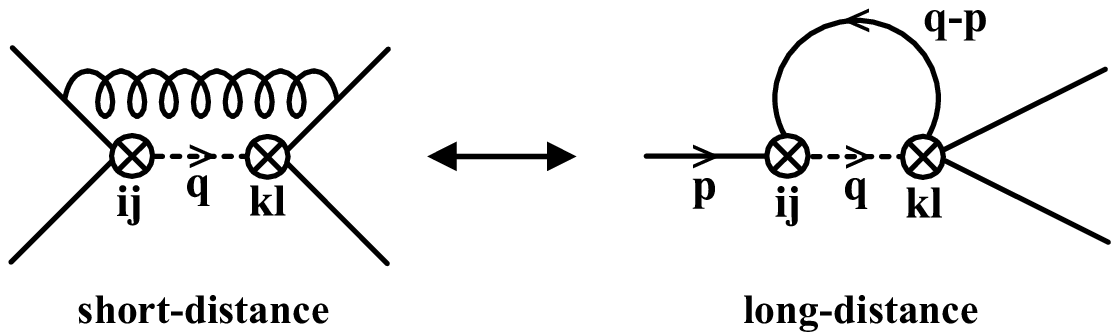,height=2.8cm}\\[12pt]
\centerline{\small Fig.~2: Matching of short- and long-distance
contributions.}\\[12pt]
Obviously the modified procedure described above is applicable only 
for the {\it non-factorizable} part of the interaction. In fig.~3 we split the
first loop diagram of fig.~1 respecting the mesonic currents which contribute 
to the matrix element $\langle \pi^+\pi^-|Q_2|K^0\rangle$, where
$Q_2\equiv(J_\mu^L)_{su}(J^\mu_L)_{ud}$.\\[12pt] 
\hspace*{2.33cm}\epsfig{file=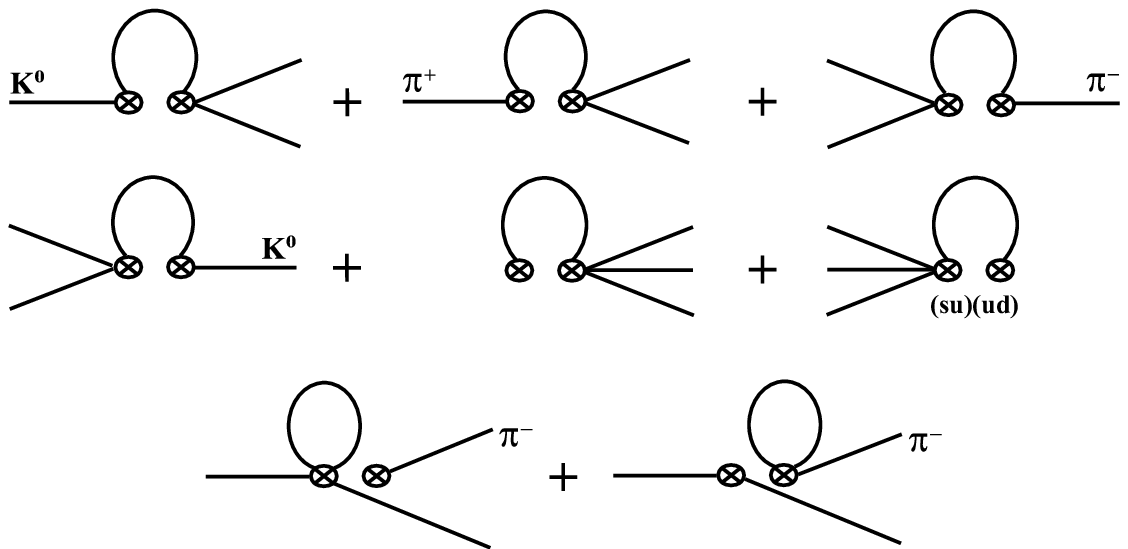,height=6.0cm}\\[12pt]
{\small Fig.~3: Representative loop contribution to the matrix element
$\langle \pi^+\pi^-|Q_2|K^0\rangle$, the cross circles denoting the currents
$(J_\mu^L)_{su}$ (left circle) and $(J^\mu_L)_{ud}$ (right) for incoming
particles.}\\[12pt]
The {\it factorizable} loop corrections shown in the third row are completely 
absorbed through the renormalization of the mesonic wave function and the bare
decay constant $f$, a pure long-distance feature connected with 
current conservation and the knowledge of the physical decay constant 
$F_{\pi}$.

Calculating all diagrams of fig.~1, splitted according to a clear 
identification of the virtual meson momenta, the evolution of the matrix 
element in the chiral limit reads  
\begin{equation}
\langle \pi^+\pi^-|Q_2(\Lambda_{cut})|K^0\rangle= F_\pi (m_K^2-m_\pi^2)
\Big[ 1+3\frac{\Lambda_{cut}^2}{(4\pi)^2}\frac{1}{f^2}+{\cal O}(1/N^2)\Big]\;,
\label{cc}
\end{equation}
the factor of 3 in front of the cutoff dependent term to be compared with
a factor of 2 obtained in ref.~\cite{BBG}. The result presented in
eq.~(\ref{cc}) is identical to the one derived from a study of the  
operator evolution \cite{FG}, where the analysis of the matrix element allows 
us to include in addition the logarithmic cutoff dependence arising through 
non-vanishing quark masses (currently being under investigation).\\[12pt]
As illustrated above a rigorous analysis of current-current hadronic
operators using the modified $1/N$ approach is feasible. The treatment of
{\bf density-density operators}, $Q_6$ and $Q_8$ being dominant for 
$\varepsilon'/\varepsilon$, requires additional remarks.

Whereas the {\it factorizable} part of the loop diagrams can be removed
in the case of current-current operators, the corresponding contributions 
have to be considered explicitly for density-density operators; 
a feature due to the absence of a conservation law for densities and, in 
addition, to the ignorance of a low-energy constant, analogous to the decay 
constant $F_\pi$ in case of the currents, in which the loop corrections 
could be absorbed.

Since a clear identification of the virtual meson momenta from a comparison
with the short-distance part is not at hand for factorizable diagrams
(representing pure long-distance phenomena), there is no prescription to 
fix a possible shift $(q\rightarrow q+p)$. Consequently, an {\it ambiguity
in the sub-leading divergent terms} occurs.\\[12pt]
On account of chiral symmetry no quartic dependence on the cutoff is present
in the matrix elements. Nevertheless, single loop diagrams may produce 
quartic terms which cancel in the sum. Thus a dependence on the (arbitrary) 
momentum shift may result in the ${\cal O}(\Lambda_{cut}^2)$. 
A concrete analysis however shows that in case of the density-density 
operators each factorizable diagram 
is at most quadratically divergent, the ambiguity only arising in the
logarithmic terms. This characteristic is based on the symmetric structure of 
the meson density given in eq.~(\ref{CD}), through which the weak vertex 
solely involves the {\it sum} of the two (incoming) virtual momenta 
$q_1$ and $q_2$,
\begin{equation}
(D_L)_{ij}\rightarrow (\partial^2 U)_{ji}\rightarrow (\partial^2 [\pi^a
\lambda_a]^n)_{ji}\rightarrow (q_1+q_2+\ldots)^2=p_{ext}^2\;,
\end{equation} 
$q_1+q_2$ being independent of the integration variable $q$ ($n$ denoting 
the number of particles, $p_{ext}$ the external momentum associated with the 
density).\\[12pt]
A strict analysis of the {\it non-factorizable} contributions can be 
performed for density-density in the same way as for current-current 
operators. Consequently, from the sum of the factorizable and the 
non-factorizable parts we can clearly determine the quadratic cutoff
dependence of the matrix elements.

Splitting the basic diagrams of fig.~1 for the gluon penguin operator
$Q_6=-2(D_RD_L)_{sd}$ we may calculate the evolution of the corresponding 
matrix elements for $K\rightarrow \pi\pi$ decays. The final result (obtained
from about 100 explicit diagrams) reads
\begin{equation} 
\langle \pi^+\pi^-|Q_6|K^0\rangle\,=\, 
\langle \pi^0\pi^0|Q_6|K^0\rangle\,=\, 
\D -F_\pi(m_K^2-m_\pi^2)\frac{r^2}{\Lambda_\chi^2}
\Big[1+3\frac{\Lambda_{cut}^2}{(4\pi)^2}\frac{1}{f^2}
+{\cal O}(1/N^2)\Big]\;,
\label{q6}
\end{equation}
where
\begin{equation}
F_\pi = f\Big[1-3\frac{\Lambda_{cut}^2}{(4\pi f)^2}\Big]\,,\hspace{1.5cm}
r^2 = \frac{(2 m_K^2-m_\pi^2)^2}{m_s^2(\Lambda_{cut})}\;.
\end{equation}
\noindent
Note that (due to the inclusion of the $\eta_0$) the parameter $r^2$ has no 
quadratic cutoff dependence.\\[12pt]
It is known that the value of the chiral symmetry breaking 
parameter $\Lambda_{\chi}$ increases through logarithmic corrections
from $\Lambda_{\chi}\mbox{(LO)}\simeq 1.02\,GeV$ to $\Lambda_{\chi}
\mbox{(NLO)}\simeq 1.20\,GeV$ (see BBG \cite{BBG}), thus 
counteracting the increase caused by the explicit loop contributions given
in eq.~(\ref{q6}). 
Consequently, we expect the $1/N$ corrections to the matrix element of $Q_6$ 
to be small.  

To perform however a detailed numerical analysis in view of the
ratio $\varepsilon'/\varepsilon$, we have to include in addition the 
logarithmic terms which arise from the {\it explicit} loop corrections. 
These terms can be calculated in a straightforward way
only for the non-factorizable diagrams. To determine the logarithmic
behaviour in the factorizable part one has to introduce a clear
prescription to fix the virtual meson momenta. As we have seen this is not
possible from a direct matching with the short-distance 
contributions. This issue is still to be investigated.
\section{Operator Evolution}
To gain a deeper understanding of long-distance effects we may study the
$1/N$ corrections to density-density operators themselves rather than those to 
the matrix elements. For that we may apply the {\bf background field method} as 
used by Fatelo and G{\'e}rard in the case of current-current operators 
\cite{FG}. This approach is powerful as
\begin{itemize}
\item the one-loop effects are given in terms of operators, too.
Therefore the analysis is {\it process independent}
(respecting the various decay channels of $K\rightarrow\pi\pi$).
\item the calculation keeps track of the {\it chiral structure} in the
loop corrections.
\end{itemize}
Whereas the method is particularly powerful in the chiral limit on account
of the (relatively) small number of diagrams which have to be calculated,
the inclusion of mass terms is difficult, wherefore we restrict our analysis to 
the case of massless quarks, i.e., to the quadratic cutoff dependence.

Starting from the chiral limit $(m_q=0)$, in the lagrangian we have to keep 
a mass term for the singlet pseudoscalar $\eta_0$ arising through the 
explicit breaking of the $U_A(1)$ symmetry,
\begin{eqnarray}
{\cal L}&=&\frac{f^2}{8}\Big( \mbox{Tr}(\partial_\mu U 
\partial^{\mu}U^\dagger)+\frac{m_0^2}{12}[\mbox{Tr}(\ln U-\ln U^\dagger)]^2\Big)
\nonumber \\
&=&\bar{\cal L}
+\frac{1}{2}(\partial_\mu\xi^a\partial^\mu\xi^a)+
\frac{1}{2}\mbox{Tr}([\partial_\mu\xi,\,\xi]\partial^\mu
\bar{U}\bar{U}^\dagger)-\frac{1}{2}m_0^2\xi^0\xi^0+{\cal O}(\xi^3)\;.
\label{la2}
\end{eqnarray}
To obtain the second line of eq.~(\ref{la2}) we decomposed the matrix $U$ in 
the quantum field $\xi$ and the classical field $\bar{U}$,   
\begin{equation}
U=\exp (2i\xi/f)\,\bar{U}\;,
\hspace{0.5cm}\xi=\lambda_a\xi^a\;,
\end{equation}
$\bar{U}$ satisfying the equation of motion
\begin{equation}
\bar{U}\partial^2\bar{U}^\dagger-\partial^2 \bar{U} \bar{U}^\dagger=
\frac{m_0^2}{N}\mbox{Tr}(\ln\bar{U}-\ln\bar{U}^\dagger)\cdot 1\;,
\hspace{0.5cm}
\bar{U}=\exp(2i\pi^a\lambda_a/f)\;.
\end{equation}
If in the same way we expand the meson density, given in eq.~(\ref{CD}), 
around the classical field (with $U\leftrightarrow U^\dagger$ for 
$L\leftrightarrow R$), which yields
\begin{eqnarray}
(D_R)_{ij}&=&(\bar{D}_R)_{ij}+\frac{if}{2}\Big(\bar{U}^\dagger\xi-\frac{1}
{\Lambda_\chi^2}
(\partial^2\bar{U}^\dagger\xi+2\partial_\mu\bar{U}^\dagger\partial^\mu\xi
+\bar{U}^\dagger\partial^2\xi)\Big)_{ji}+\frac{r}{2}\Big[\bar{U}^\dagger\xi^2
\label{Dexp}\\
&&\hspace{0mm}-\frac{1}{\Lambda_\chi^2}\Big(\partial^2
\bar{U}^\dagger\xi^2+2\partial_\mu\bar{U}^\dagger\{\partial^\mu\xi,\,\xi\}
+\bar{U}^\dagger
(\{\partial^2\xi,\,\xi\}+2\partial_\mu\xi\partial^\mu\xi)\Big)\Big]_{ji}
+{\cal O}(\xi^3)\;,\nonumber
\end{eqnarray}
we may calculate the $1/N$ correction to a density-density
operator of the form $(D_R)_{ij}(D_L)_{kl}$ by integrating out the
$\xi$ field, the terms linear in $\xi$ in eq.~(\ref{Dexp}) contributing to the
{\it non-factorizable} one-loop diagrams, the quadratic to the 
{\it factorizable} ones, where the rescattering of the mesons due to the
strong interaction may be included using the lagrangian (\ref{la2}).
The external legs of a diagram (see fig.~4) now symbolize weak and
strong operators, respectively, given in terms of the field $\bar{U}$.
Consequently, the loop corrections preserve the chiral structure.\\[12pt]
\hspace*{3cm}\epsfig{file=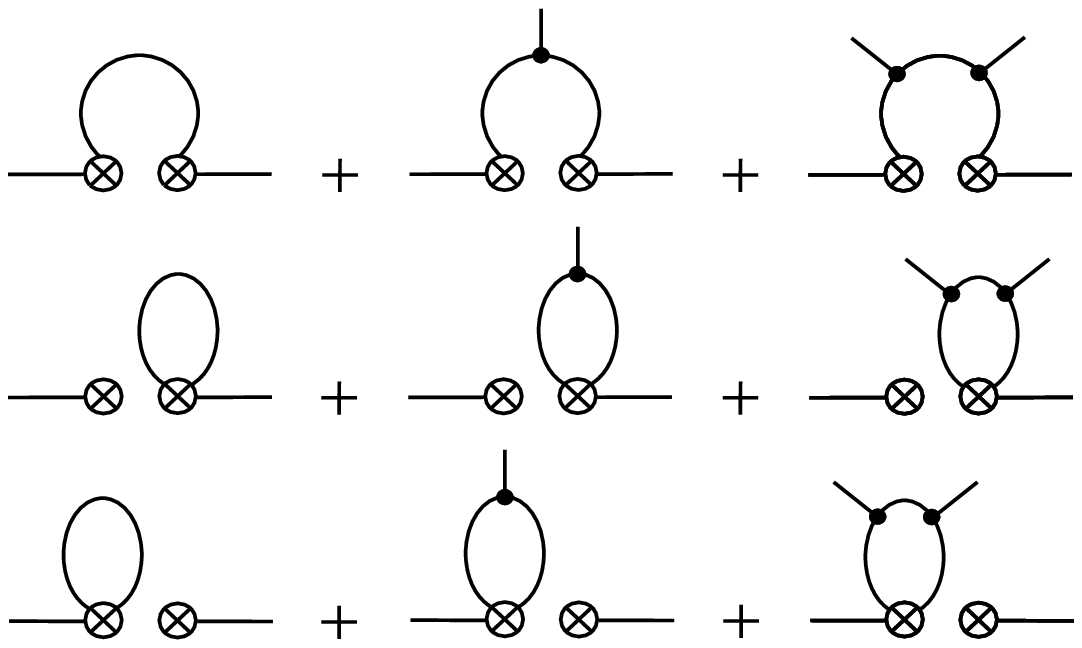,height=6.2cm}\\[12pt]
{\small Fig.~4: One-loop ${\cal O}(p^2)$ contribution to the long-distance 
evolution of the operator $(D_R)_{ij}(D_L)_{kl}$, the cross circles denoting 
the densities, the black circles the strong interaction.}\\[12pt]
The {\it non-factorizable} part (first row of fig.~4) may again be calculated 
in a straightforward (although lengthy) way associating the cutoff 
to the effective color singlet gauge boson connecting the two 
densities. The {\it factorizable} contributions however (second and third
row of fig.~4), involving
trilinear divergences through a certain class of diagrams, show a dependence 
on a shift of the loop momenta ($q\rightarrow q+p$) already in the quadratic 
cutoff behaviour. This feature, occuring in the study of the operator 
evolution by means of the background field approach, does not appear in the 
explicit analysis of the various matrix elements 
(see section \ref{CLC}).\\[12pt]
In order to avoid an ambiguity of the operator evolution in the chiral
limit, it is sufficient to adopt a prescription for the {\it distribution} of 
momenta in the concerned diagrams. Indeed we have strong indications for a 
particular choice (based on arguments beyond the scope of this
talk), which removes any further shift dependence from the quadratic terms.   
This choice is natural but not necessarily unique.

Thus performing the analysis, from the sum of the factorizable and the 
non-factorizable parts we obtain the long-distance evolution of the operator 
$Q_6$,
\begin{equation}
Q_6(\Lambda_{cut}^2)=-\frac{F_\pi^4}{4}\frac{r^2}{\Lambda_\chi^2}
(\partial_\mu \bar{U}\partial^\mu \bar{U}^\dagger)_{ds}(0)
\Big[1+3\frac{\Lambda_{cut}^2}{(4\pi f)^2}+{\cal O}(1/N^2)\Big]\;,
\end{equation}
the chiral loop corrections being identical to those found in the study
of the matrix element (using the background field approach the
renormalization factors for the bare decay constant $f$ and the bare meson 
field $\pi^a$ cancel in the matrix $\bar{U}$).

Note that the result is given in terms of $(\partial_\mu\bar{U}\partial^\mu
\bar{U}^\dagger)_{ds}$, the latter being the only pseudoscalar operator 
available at the ${\cal O}(p^2)$ to describe both, the corresponding 
current-current {\it and} density-density four-fermion operators
($Q_4$ and $Q_6$, respectively).
\section{Summary and Conclusion} 
We studied the one-loop long-distance contributions to $K\rightarrow
\pi\pi$ decay amplitudes applying the $1/N$ expansion in which we used the
momentum flow between the currents and densities, respectively, to fix the 
virtual meson momenta. Thus a proper matching with the short-distance part
is feasible for non-factorizable diagrams.

The matrix elements of density-density operators, containing factorizable 
contributions which have to be considered explicitly, were found to be 
clearly determined regarding the quadratic cutoff behaviour, yet leaving us
with an ambiguity in the logarithmic one.  
In the evolution of the operators themselves, which we investigated by means 
of the background field approach, the corresponding ambiguity already appeared 
in the quadratic divergences. However, using a prescription motivated by
theoretical arguments the methods are in agreement.

As a first analytic result we presented the $1/N$ corrections to the
gluon penguin operator $Q_6$ as well as to the corresponding matrix
elements in the chiral limit. The corrections are small, provided the
change in the value of the hadronic scale $\Lambda_{\chi}$ is considered.
A detailed numerical analysis in view of the ratio $\varepsilon
'/\varepsilon$  requires in addition the inclusion of logarithmically 
divergent terms.This issue is still under investigation.\\
%%%%%%%%%%%%%%%%%%%%%%%%%%%%%%%%%%%%%%%%%%%%%%%%%%%
\vspace{3cm}\\
\noindent
\centerline{ \large \bf  Acknowledgements }
The analysis presented in this talk is based on a collaboration with
Th.~Hambye, G. K\"ohler and E.A. Paschos.

PHS wants to thank the {\it Deutsche Forschungsgemeinschaft}
for financial support (in connection with the Graduate College for
Elementary Particle Physics in Dortmund).
Special thanks are due to the organizers of the workshop for a very nice 
and instructive meeting.
\newpage
%

%%%%%%%%%%%%%%%%%%%%%%%%%%%%%%%%%%%%%%%%%%%%%%%%%%%%%%%%%%
\end{document}